\documentclass[aps,superscriptaddress,pre,showpacs,twocolumn]{revtex4}
\usepackage{graphicx,amssymb,amsmath,subeqnarray}
\newcommand{\m}{{\bf m}}
\newcommand{\rr}{{\bf r}}
\def\d{{\rm d}}
\newcommand{\dt}{\partial_t}
\newcommand{\dl}{\delta\ell}

\def\bmu{{\boldsymbol \mu}}
\newcommand{\muinv}{\mu^{-1}}

\begin{document}
\title{Shaking-induced motility in suspensions of soft active particles}
\author{Denis Bartolo}
\email{denis.bartolo@espci.fr}
\affiliation{Laboratoire de Physique et M\'ecanique des milieux h\'et\'erog\'enes, CNRS,ESPCI, Universit\'e Paris 6, Universit\'e Paris 7}
\affiliation{Laboratoire Jean Perrin, CNRS, Universit\'e Paris 6.}
\author{Eric Lauga}
\email{elauga@ucsd.edu}
\affiliation{Department of Mechanical and Aerospace Engineering, 
University of California San Diego, 9500 Gilman Dr., La Jolla CA 92093-0411, USA.}

\begin{abstract} 
We investigate theoretically the collective dynamics of soft active particles living in a viscous fluid. We focus on a minimal model for active but non-motile particles consisting of $N>1$ elastic dimers  deformed by active stresses and interacting hydrodynamically.  We first derive a set of effective equations of motion for the positions of the particles. We then exploit these equations in two experimentally-relevant cases:  uncorrelated random internal stresses, and uniform monochromatic external shaking. In both cases, we show that small groups of intrinsically non-motile particles can display non-trivial modes of locomotion resulting from the hydrodynamic correlations between the particle-conformation fluctuations. 
In addition, we demonstrate that a coherent shaking yields spatial ordering in suspension of  soft particles interacting solely through the fluid.

\end{abstract}
\maketitle


\section{Introduction}
Active suspensions generically refer to an ensemble of particles which continuously convert energy supplied by internal or external sources to deform and/or propel themselves in a fluid. Classical examples include suspensions of biological swimmers  such as bacteria, algae or spermatozoa~\cite{kessler97,mendelson99,libchaber00,dombrowski04}, microscopic robotic devices~\cite{sen09,wang09,ghosh09}, and field-responsive particles which change their shape in response to external applied fields (such as electric, magnetic or  temperature fields...)~\cite{rayleigh1882,goubault03,buguin06}. 

Recently, a number of theoretical studies have been devoted to understand the large scale properties of this important class of non-equilibrium systems ~\cite{pedley92,simha02,Liverpool03,toner05,Kruse05,saintillan08}. Most of these coarse-grained models  describe the large scales features of the active media, and ignore the specifics of the deformations or propulsion mechanisms at the single particle level. For instance, to model a suspension of swimming bacteria, the deformation pattern of the bacterias body is not explicitly described, but instead the cells are modeled as  constant, local, and anisotropic stress distributions acting on the surrounding fluid~\cite{batchelor70_2,pedley92}. 
The  argument used to  discard the details of the deformation dynamics   is that these hydrodynamic models should be valid  for time scales larger than the typical time-scale of the periodic microscopic actuation. 

However, this implicit time averaging step is not trivial, and it may in fact hide some subtle collective effects~\cite{koiller96,ishikawa07_bacteria,alexander08,laugabartolo08}. Indeed, {a priori}, for a given active internal stress, the instantaneous rate of deformation of a soft active particle depends  on the conformation, the orientation, and  the position of all the other particles due to long-ranged hydrodynamic interactions. This leads to a  natural question to ask: Can the hydrodynamic coupling between the unsteady conformation and the particles positions  yield unexpected phase behavior and new modes of locomotion in suspensions of soft active particles? 

Here, we investigate a simple class of active suspensions comprising  actively-deformed, but not-self-propelled, soft particles which interact only through the surrounding fluid. We  start by rigorously deriving the effective equations of motions for the center  of the particles  in the dilute limit.
After discussing the case of deformation-controlled dynamics, we then consider  the coupled conformational and translational dynamics of the suspension.
We focus on two opposite, but experimentally relevant,  situations. First, we assume that the particles are subject to  stochastic but uncorrelated internal actuation.  This is probably the simplest description to account for the shape changes induced by conformational-transitions in nano-machines~\cite{golestanian08}, and the shape fluctuations of non-motile living cells~\cite{pullarkat07}. In that case, the center of a group of identical soft active-particles only experiences  correlated random motion, 
but suspensions  of heterogeneous particles  can display cooperative directed motion. In addition, finite-size suspensions  evaporate or condense depending on their spatial distribution.
Next, we  consider  the case where the particles are subject to a deterministic and uniform shaking at a prescribed frequency. This type of forcing is relevant to model the  actuation of soft colloids, or more generally field-responsive particles, by external fields. In that case, although isolated particles  generate no net flow,  
hydrodynamic coupling induces nontrivial collective swimming modes for either identical or non-identical particles, as well long-range  effective interactions. Unbounded suspensions including a finite number of identical particles always evaporate, while polydisperse suspensions can also condense depending on their relative spatial distribution. Remarkably, in the thermodynamic limit, systems made of identical particles display a shaking-induced crystallization.

The paper is organized as follows. We introduce our minimal model of
soft active particle suspension  in section~\ref{model}.
The general derivation of the effective equations of motion for the noise- or time- averaged positions of the $N$ particles for the two opposite shaking sources (coherent and incoherent) is then offered in section~\ref{derivation}. The case of incoherent (stochastic) forcing is studied in detail in section~\ref{incoherent}, while the case of coherent (deterministic) forcing is addressed in section~\ref{coherent}. We finish by some concluding remarks, and some suggestions for future work. 
\section{Model}\label{model}
\begin{figure}
  \begin{center}
    \includegraphics[width=0.95\columnwidth]{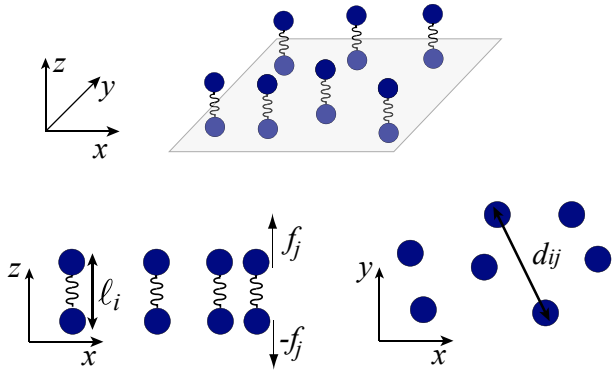}
    \caption{Minimal model of soft active particle suspensions: 
    $N$ elastic dimers moving in the $(x,y)$ plane. Top: 3D view. Bottom left: side view. Bottom right: top view.}
    \label{geometry}
  \end{center}
\end{figure}

Our minimal model is illustrated in Fig.~\ref{geometry}. 
We consider $N$ dimers (what we will refer to as ``particles''), each of them being made  of two spheres of radius $a$ connected by frictionless harmonic springs. We note $k_i$ and and $\ell_i(t)\equiv\ell+\delta\ell_i(t)$ the stiffness and the length of the $i^{\rm th}$ dimer. All the particles have the same shape but can have different stiffnesses. The dimers live in a three-dimensional viscous fluid (Stokes flow), but are assumed for simplicity to remain in the two-dimensional $(x,y)$ plane, and  oriented along the $z$-axis, see Fig.~\ref{geometry}. 
In that case, the dynamics of the system is described by the elongation $\ell_i$ and of the position $\rr_i$ of each of the $N$ dimers. Each  dimer in the suspension is force-free, but  can expand and contract due to internal stresses modeled here by two equal and opposite forces acting on each of the two spheres. These forces of amplitude $f^{\rm act}_i(t)$ are assumed to be oriented along the dimer axis, and the particles are thus deformed only in the direction normal to their motion.  The linearity of the stokes equation implies that the velocities and the elongation rates of the dimers are linearly related to the total internal forces, which are the sum of an elastic (restoring) and of an active components, $f_i(t)=f^{\rm el}_i(t)+f^{\rm act}_i(t)$, and thus
\begin{eqnarray}
\dt\rr_i&=&\sum_j\m_{ij}\left(\lbrace\ell_i\rbrace,\lbrace\rr_{ij}\rbrace\right)f_j \label{baremotion},\\
\dt\ell_i&=&\sum_j\mu_{ij}\left (\{\ell_i\},\{\rr_{ij}\}\right)f_j.\label{conf}
\label{bareconformation}
\end{eqnarray}
In Eqs.~\eqref{baremotion} and \eqref{bareconformation}, $f_j $ refers the $z$-magnitude of the force acting on the top dimer in Fig.~\ref{geometry} (bottom left). In principle all the mobility coefficients $m_{ij}$ and $\mu_{ij}$ in Eqs.~\eqref{baremotion} and \eqref{bareconformation}  depend both on the elongation of all the particles, as well as on all the inter-particles distances, $r_{ij}(t)\equiv d_{ij}+\delta d_{ji}(t)$. This naturally makes the above equations highly nonlinear, and in general impossible to solve by hand. In order to obtain analytical results we restrict our study to the limit of dilute suspensions, $d_{ij}/\ell_i\gg1$, and  wide dimers, $\ell/a\gg1$. In that limit, the $2N$ spheres composing the $N$ dimers can be replaced by point forces  (Stokeslets). The velocity of the $n^{\rm th}$ sphere is thus related to the force acting on the $m^{\rm th}$ sphere by $\partial_t {\bf R}_n=\sum_m{\bf G}({\bf R}_n-{\bf R}_m)\cdot {\bf F}_m(t)$, where ${\bf R}_n$ is the position of sphere $n$, and ${\bf G}=\left({\bf 1}+{{\bf RR}}/{R^2}\right)/(8\pi\eta R)$ is the Oseen kernel~\cite{happelbrenner}. Within this approximation, the mobility coefficients  defined in Eqs.\eqref{baremotion} and \eqref{bareconformation} can be computed analytically  at the necessary order in  in $a/\ell$ and  $\ell/r$. Setting $G(r)\equiv1/(4\pi\eta r)$, and $G(0)\equiv1/(6\pi\eta a)$, it is straightforward to get 
\begin{eqnarray}
&&\mu_{ii}=2G(0)\equiv\mu_0,\label{muii}\\
&&\mu_{ij}=-\frac{1}{4}\ell_i\ell_jG''(r_{ij})\label{muij}, \, i\neq j\\
&&\m_{ii}=0\label{mii},\\
&&\m_{ij}={\bf e}_{ij}\left[- \frac{\ell_j}{2}G'\left(r_{ij}\right)+\frac{\ell_j(3\ell_i^2+\ell_j^2)}{32}G'''(r_{ij})\right]\label{mij}, \, i\neq j \quad \,\,
\end{eqnarray}
where ${\bf e}_{ij}$ is the unit vector pointing from the center of mass of dimer $i$ to that of dimer $j$, and with higher-order terms which are negligible in what follows.  Importantly, we note that $\bmu$ is symmetric and that $\bf m$ is not antisymmetric. At higher order in the separation distance between the dimers, only corrections with even (resp. odd) derivative in $G$ add up to the $\mu_{ij}$ (resp. to the  $\m_{ij}$). 
Finally, we note  from Eqs.~\eqref{baremotion} and \eqref{mii} that the  position of an isolated dimer does not vary in time: The particles are therefore active, but are not self-propelled. This result is a direct consequence of Purcell's  scallop theorem  stating that body deforming with time-reversible strokes cannot display any average motion in a Stokes flow \cite{purcell77}.

\section{Effective dynamics}
\label{derivation}
In this section, we derive the coarse-grained equations of motion for the dimers. Specifically, we explain how to adequately perform averages for  Eq.~\eqref{baremotion}. 
All along this section, we use the symbol $\langle ... \rangle $ to denote averages. Anticipating the two types of actuation of  interest, these averages can represent two different things. In the case of a stochastic forcing, the  averages  are made over a large number of noise realizations (section \ref{incoherent}). In the case of a deterministic AC forcing, they refer to time-averages over the actuation period  (section \ref{coherent}). The algebra in both cases is identical, so we present it below without specifying which of the situation we are referring to. The final equation  we obtain is Eq.~\eqref{effective}, and the reader uninterested in technical details can skip what precedes it.

\subsection{Derivation of the coarse-grained equations of motion}
We restrict ourselves to the limit of small deformations,  $\delta \ell_i/\ell_i\ll1$ and small displacements $\delta d_{ij}/d_{ij}\ll1$. Expanding the mobility coefficients, $\m_{ij}$ at first order in $\delta \ell$ and averaging Eq.~\eqref{baremotion} yields
\begin{eqnarray}
&&\langle\dt\rr_i\rangle=\label{dtriordre2}\\
&&\sum_j\left [\underbrace{\m_{ij}\langle f_j\rangle}_{{\bf  A_{ij}}}+\underbrace{\frac{\partial \m_{ij}}{\partial d_{ij}}\langle \delta d_{ij}f_j\rangle}_{{\bf  B_{ij}}}
+\underbrace{\sum_{n=i,j}\frac{\partial \m_{ij}}{\partial \ell_{n}}\langle \delta \ell_{n}f_j\rangle}_{{\bf  C_{ij}}}\right ]\nonumber,
\end{eqnarray}
where Taylor expansions have been used, and therefore all the $\m_{ij}$ terms in 
Eq.~\eqref{dtriordre2} are evaluated at $\ell_i$ and $d_{ij}$. Importantly, in Eq.~\eqref{dtriordre2} and throughout the paper, we will not be using summation over repeated indices, and therefore sums will be explicitly written down.

We are now going to show that in the dilute limit, the right hand side of Eq.~\eqref{dtriordre2} is dominated by the ${C}_{ij}$ terms. To identify this dominant contribution, we expand the forces from Eq.~\eqref{bareconformation} at second order in the $\delta \ell_i$ and $\delta d_{ij}$ and obtain
\begin{eqnarray}
f_j=\sum_k\left(\muinv_{jk}\delta\dot{\ell}_k\!+\!\frac{\partial\muinv_{jk}}{\partial d_{jk}}\delta d_{jk}\delta\dot{\ell}_k\!+\!\!\!\sum_{n=j,k}\!\!\frac{\partial\muinv_{jk}}{\partial \ell_{n}}\delta \ell_{n}\delta\dot{\ell}_k\right), \label{fiordre2}
\end{eqnarray}
where $\muinv_{mn}$ refers to the $mn$ component of the matrix $\boldsymbol{\mu}^{-1}$ inverse of $\bmu$. We then use this expansion to compute the ${A}_{ij}$, ${ B}_{ij}$ and ${ C}_{ij}$ terms in Eq.~\eqref{dtriordre2}. 

To do so, we fist need to compute the inverse of the mobility matrix $\bmu$. In the dilute limit, we get from  Eqs.~\eqref{muii} and~\eqref{muij} that $\mu_{ii}=\mu_0\gg\mu_{ij}$ for all $i$ and $j\neq i$.  As a result, we  can expand $\bmu$ at first order in the $\mu_{ij}/\mu_0\sim \ell^2G''(d_{ij})/G(0)$, and invert the matrix to obtain at the same order
\begin{eqnarray}
\mu^{-1}_{ii}&=&\frac{1}{\mu_0},\label{muinvii}\\
\mu^{-1}_{ij}&=&-\frac{\mu_{ij}}{\mu_0^2}\label{muinvij}, \quad i\neq j.
\end{eqnarray}

Let us now first consider the order of the terms ${\bf A}_{ij}$ and ${\bf B}_{ij}$.  
At lowest order in $\delta \ell/d$, $\delta d/d$ and $a/\ell$ we have 
\begin{equation}\label{A}
{\bf A}_{ij}=
\m_{ij}\sum_{k}\frac{\partial\muinv_{jk}}{\partial d_{jk}}\langle\delta d_{jk}\delta\dot{\ell}_k\rangle
+
\m_{ij}\sum_{k\neq j}\frac{\partial\muinv_{jk}}{\partial \ell_{j}}\langle\delta \ell_{j}\delta\dot{\ell}_k\rangle,
\end{equation}
and
\begin{equation}\label{B}
{\bf B}_{ij}=\sum_k \frac{\partial \m_{ij}}{\partial d_{ij}}\muinv_{jk}\langle \delta d_{ij}\delta\dot{\ell}_k\rangle,
\end{equation}
where we have used $\langle \delta \dot\ell_n\rangle =0$ and $\langle \delta\ell _i\delta\dot\ell_i\rangle =0$. Since we have $d_{ij}^2 = ({\bf r}_j-{\bf r}_i)^2$, we can  use Eqs.~\eqref{dtriordre2} and \eqref{fiordre2} to write $\delta d_{ij} = {\bf e}_{ij} \cdot (\delta {\bf r}_j- \delta {\bf r}_i) $ at first order in the elongation variables, and therefore get
\begin{equation}
\delta d_{ij}={\bf e}_{ij}\cdot \left[\sum_{n,m}\left(\m_{jn}-\m_{in}\right)\muinv_{nm}\delta \ell_m\right]\cdot
\label{deltaddeltal}
\end{equation}
Using Eq.~\eqref{deltaddeltal} in Eqs.~\eqref{A} and \eqref{B}, and recognizing again that 
 $\langle \delta \ell_i \delta \dot{\ell}_i\rangle=0$, we see that the coefficients in front of averages of the form $\langle \delta \ell_j \delta \dot{\ell}_k\rangle$ in both ${\bf A}_{ij}$ and ${\bf B}_{ij}$ are of the order of $\m$ times an off-diagonal term of the matrix $\bmu^{-1}$, and therefore decay in space as ${\cal O}(G'G'')\sim  1/r^5$ for  an unbounded fluid.

We now turn to the  ${\bf C}_{ij}$ term. At lowest order we have
\begin{eqnarray}
{\bf C}_{ij}&=&\mu_0^{-1}\frac{\partial \m_{ij}}{\partial \ell_i}\langle\delta \ell_i\delta \dot{\ell}_j\rangle\\
&=&\frac{3}{16\mu_0}\ell^2G'''(d_{ij})\langle\delta \ell_i\delta \dot{\ell}_j\rangle{\bf e}_{ij}.
\end{eqnarray}
Consequently, in the  ${\bf C}_{ij}$ term, the coefficient in front of the conformation averages are on the order of $G'''$, which decays in space as $\sim 1/r^4$ in an unbounded fluid. Since $G''' \gg G'G''$, we see  therefore that, in the dilute limit, the ${\bf C}_{ij}$ term in Eq.~\eqref{dtriordre2} dominates the dynamics of the time-averaged positions of the dimers. Consequently, 
Eq.~\eqref{dtriordre2}  reduces to the final coarse-grained equation
\begin{equation}
\partial_t\langle\rr_i\rangle=\frac{3 \ell^2}{16\mu_0}\sum_{j\neq i}G'''(d_{ij})\langle\delta \ell_i\delta\dot{\ell}_j\rangle \,{\bf e}_{ij},
\label{effective}
\end{equation}
which is our first main result.
\subsection{Deformation-controlled scenario}
For a system composed of more than one particle, Eq.~\eqref{effective} shows that the dimers positions will  evolve  in time if their conformations have nontrivial correlations. So, a collection of non-swimmers can be made to swim. Physically, this cooperative motion arises due to the coupling between the location of each particle and the instantaneous dipolar flow field induced by the deformations of the surrounding particles. 

In the case where the conformation kinematics is prescribed, then Eq.~\eqref{effective} fully prescribes the suspension dynamics. For instance, let us consider the simple case of two identical dimers with sinusoidal length modulations: $\ell_i(t)=\ell+\delta \ell \cos(\omega t+\phi_i)$. It is then straightforward to derive from 
Eq.~\eqref{effective}  the  overall average swimming velocity, ${\bf v}\equiv\frac{1}{2}\langle\partial_t(\rr_1+\rr_2)\rangle$, and overall relative velocity,  $\Delta {\bf v} \equiv\partial_t(\rr_2-\rr_1) $, and  we find
\begin{eqnarray}
{\bf v}&=&\frac{27}{64}\left(\frac{\ell\delta \ell}{d^2}\right)^2 a\omega \sin(\phi_{2}-\phi_1)\,{\bf e} _{12},
\label{v}\\
\Delta {\bf v} & = & {\bf 0}.
\end{eqnarray}
The two dimers remain at a constant distance on average.
If $\phi_2-\phi_1= 0$ or  $\pi$, then the deformations of the two-particle system are reciprocal,  and average swimming is prohibited by the scallop theorem. In all other cases, the two particles  swim collectively. There is therefore no many-scallop theorem~\cite{laugabartolo08,alexander08}. 

Next, and more formally,  it is important to recognize that the effective dynamics of $N$ soft active particles cannot be mapped on the relaxation dynamics of an ensemble of passive particles interacting through a long-ranged pair potential. This is because the  conformation correlation terms in Eq.~\eqref{effective} has the properties of  a pseudo-scalar. Indeed, the length fluctuations are either periodic or statistically stationary, which implies that $\langle\delta \ell_i(t)\delta\dot\ell_j(t)\rangle=-\langle\delta \ell_j(t)\delta\dot\ell_i(t)\rangle$. In addition, $G'''(d_{ij}){\bf e}_{ij}\sim\nabla G''(d_{ij})$ derives from a scalar quantity, this implies that the full term in Eq.~\eqref{effective} cannot derive from a scalar potential, and our conclusion is therefore true for all type of actuation applied to the soft particles.
\label{pseudo}

In the sections below we focus on  two different, and more complex, types of actuation. 
First we consider the case of particles shaken by internal stochastic forces having no-spatial and no-temporal correlations (incoherent shaking, section~\ref{incoherent}). Next we  study the case of a uniform sinusoidal external force acting on all the soft particles (coherent shaking, section~\ref{coherent}).  In both cases, we will solve Eq.~\eqref{conf} whithin a linear response approximation to compute the conformation correlations, $\langle\delta \ell_i \delta\dot\ell_j \rangle$, and  Eq.~\eqref{effective} will then be used to derive the effective dynamics of the suspension.
\section{Incoherent shaking}\label{incoherent}
\subsection{Effective dynamics}
In this first scenario, we assume that the active component of the force on each sphere in a dimer is a random variable with zero mean and fluctuations given by
\begin{eqnarray}
\langle f^{\rm act}_i(t)\rangle&=&0,\\
\langle f^{\rm act}_i(t)f^{\rm act}_j(t')\rangle&=&f^2\tau\delta(t-t')\delta_{ij},\label{noise}
\end{eqnarray}
where $f$ is the typical force amplitude and $\tau^{-1}$ the highest frequency  of the force power spectrum.  Since $f^{\rm act}$ is not spatially correlated whereas the mobility matrix $\bmu$  is non-local (see Eq.~\ref{muij}), the fluctuation-dissipation relation cannot for this noise term, and therefore $f^{\rm act}$ does not represent thermal noise. 
Instead, the term $f^{\rm act}$ models active and fluctuating small-scale forcing, for example the stochastic actuation by  molecular motors.

To derive the conformation correlations we separate explicitly the active component of the force from its elastic component in Eq.~\eqref{conf}
\begin{equation}
\partial_t \delta \ell_i=-\sum_j \mu_{ij}k_j\delta\ell_j + \mu_{ij}f^{\rm act}_j(t),
\label{eq1bruit}
\end{equation}
where we evaluate the $\mu_{ij}(\ell_i,\ell_j,r_{ij})$  term at $(\ell,\ell, d_{ij})$. Within this linear response framework, we Fourier transform
Eq.~\eqref{eq1bruit} to obtain
\begin{equation}\label{Flin}
\delta \tilde \ell_i(\omega)=\sum_jR_{ij}(\omega) \tilde f_j(\omega),
\end{equation}
where ${\bf R}(\omega)\equiv\left[i\omega{\bf I}+\bmu{\bf k}(\omega)\right]^{-1}\bmu(\omega)$ is the kernel given by Eq.~\eqref{eq1bruit}, and the matrix $\bf k$ is defined by $k_{ij}=k_i \delta_{ij}$.

By writing the Fourier-transform of Eq.~\eqref{noise} as $\langle \tilde f^{\rm act}_i(\omega) \tilde f^{\rm act}_j(\omega')\rangle=2\pi f^2\tau\delta(\omega + \omega')\delta_{ij}$, we can then use Eq.~\eqref{Flin} to obtain the correlation as given by
\begin{eqnarray}\label{corrl}
&& \langle\delta \ell_i(t)\delta\dot{\ell}_j(t')\rangle=\\
&&-\frac{f^2\tau}{2\pi }\int\sum_k i\omega e^{i\omega(t-t')}\left[R_{ik}(\omega)R_{jk}(-\omega)\right]\notag  \d\omega
.\end{eqnarray}
We next calculate the coefficients of the response matrix $\bf R$ for dilute suspensions. In this limit,  the off-diagonal terms of $\bmu$ are subdominant,  and after some straightforward algebra,  and at the lowest order in the $1/d_{ij}$,  we get that the $R_{ij}$ terms reduce to
\begin{eqnarray}
R_{ii}&=&\frac{\mu_0}{i\omega+\omega_i},\label{Rii}\\
R_{ij}&=&\frac{i\omega\mu_{ij}}{(i\omega+\omega_i)(i\omega+\omega_j)}\label{Rij}\quad i\neq j,
\end{eqnarray}
where we have defined the relaxation frequency of the dimers by $\omega_i\equiv \mu_0 k_i $.
The sum over $k$ in Eq.~\eqref{corrl} can next be  approximated by the terms at  the leading order in $1/d_{ij}$. The dominant contribution come from the two terms in the form of a product of diagonal terms in $\bf R$ times an off-diagonal term, {\it i.e.}
\begin{eqnarray}
&&\sum_k \left[R_{ik}(\omega)R_{jk}(-\omega)\right] = \\
&& R_{ii}(\omega)R_{ji}(-\omega)
+
R_{ij}(\omega)R_{jj}(-\omega)
+ {\rm h.o.t.}
\notag\end{eqnarray}
where the higher-order terms decay faster in space.
Using Eq.~\eqref{Rii} and \eqref{Rij}, and after integration of Eq.~\eqref{corrl} over $\omega$ we get the correlation as given by
\begin{eqnarray}\label{corr_final}
\langle\delta \ell_i(t)\delta\dot{\ell}_j(t)\rangle=\frac{1}{2}(f^2\tau)\mu_0\mu_{ij}\frac{\omega_i-\omega_j}{\omega_i+\omega_j}\cdot
\end{eqnarray}
Finally, using the definiton of the $\mu_{ij}$, Eq.~\eqref{muij}, and the above expression for the correlation in the equation of motion  Eqs.~\eqref{effective}, we  obtain the final expression for the effective equations of motions of the $N$ soft dimers as
\begin{equation}
\partial_t\langle \rr_i\rangle=\frac{3\ell^4(f^2\tau)}{128}\sum_{j\neq i}
\frac{\omega_j-\omega_i}{\omega_i+\omega_j}
G''(d_{ij})G'''(d_{ij})\,
{\bf e}_{ij}.
\label{effectivebruit1}
\end{equation}
\subsection{General comments}
The general equation describing the average motion of each soft active particle is Eq.~\eqref{effectivebruit1}. Interestingly, we first note that this final formula does not involve the conformational self-mobility coefficients of the individual particles, $\mu_0$, emphasizing the role of hydrodynamic interactions. In particular, the same final formula would be obtained for dimers of different shapes and sizes. 
The fast spatial decay of the velocities in Eq.~\eqref{vnoise} with the inter-particle distance should also be stressed. In an unbounded fluid, $G\sim1/d$ and therefore $G''G'''\sim 1/d^7$.  This is to be contrasted with the $G'''\sim 1/d^4$ power law predicted when the deformations are imposed (see Eq.~\ref{effective}).  This difference arises from the $1/d^3$ decay of the elongation-elongation correlation function in a force controlled set-up, whereas the correlation is by construction independent of the distance when the deformations are imposed.
\subsection{Small groups of soft active particles: $N=2$}
The right-hand side of the equations of motion, Eq.~\eqref{effectivebruit1}, is the sum of the $N$ pairwise hydrodynamic interactions between the dimers.
Some additional insight on the collective dynamics can thus be gained by considering first the case of a single pair of dimers.
For $N=2$, it is easy to see in that  the average and  relative speeds are given by
\begin{eqnarray}
{\bf v}&=&\frac{1}{2}\left(\frac{3}{16\pi}\right)^2 \left(\frac{\omega_1-\omega_2}{\omega_1+\omega_2}\right)\frac{f^2\tau\ell^4}{\eta^2d^7}\, {\bf e}_{12},
\label{vnoise}
\\
\Delta {\bf v}&=&{\bf 0}.
\end{eqnarray}
The two particles remain at a constant distance for all noise amplitude. In fact, if the two particles are identical (same relaxation time) they do not move at all in average ($\bf v=0$). If however they posses  different relaxation times, the particles move  together with the same speed oriented along the ${\bf e}_{12}$ direction, and with  a sign that depends on the relative magnitude of their relaxation frequencies. This collective mode of locomotion arises despite the fact that (i) the two particles are non-motile when isolated and (ii) the stochastic shaking forces do not induce any  averaged fluid velocity in the far field.

\subsection{Interpretation}

To understand intuitively the existence of a net locomotion in the presence of noise, we can think of the stochastic shaking forces as the incoherent superposition of sinusoidal forces with uncorrelated random phases. The response to two forces of the form $f^{\rm act}_i=f\int d\omega \cos[\omega t+\psi_i(\omega)]$ results in out-of-phase oscillations of the two dimers lengths, with a phase difference having two contributions. The first contribution comes obviously from the phase differences between the two shaking forces; the second intrinsic contribution arises from the difference between the two relaxation times. 

Let us first ignore the off-diagonal coupling of the conformation mobility tensor $\mu_{ij}$. In that case, the dimer $i$ would respond solely to the force $f_i$, and would thus oscillate with the phase $\omega t+\psi_i+\phi_i$, where $\phi_i$ is the phase shift due to the finite response time of the elastic spring coupled to the viscous fluid. In turn, the elongation of the first dimer and the elongation rate of the second dimer would be necessarily uncorrelated as the shaking forces have uncorrelated phases. More precisely, the average velocity given by Eq.~\eqref{v} scales with $\sin(\phi_1+\psi_1-\phi_2-\psi_2)$,  which averages to zero for any two relaxation times.

If we now take into  account the off-diagonal hydrodynamic coupling, we can physically interpret the appearance of locomotion. Indeed, as momentum propagates instantaneously in the fluid (there is no inertia), the elongation of, say, the first dimer, $i=1$, has now two contributions. The first contribution is the local response to the local force $f_1$, while the second results from the dipolar flow field which oscillate in phase  with the dipolar force sources, and is induced by the force $f_2$ acting on the second dimer. The local contribution to the elongation has a phase given by $\omega t+\psi_1+\phi_1$, whereas the non-local contribution to the deformation is oscillates with $\omega t+\psi_2+\phi_1$. It then follows that the local contribution to the elongation of the first dimer correlates with the non-local contribution to the elongation rate of the second dimer if the two particles have different relaxation times. Indeed, this correlation scales now with $\sin(\phi_1+\psi_1-\phi_2-\psi_1)$ (and an amplitude decaying with $d$),  which does  not average to zero as the dimers are distinct.  
  Note that such interpretation also holds for viscoelastic dimers which internally dissipate energy (results not shown). 
\subsection{Small groups of soft active particles: $N=3$}
Before addressing the dynamics of large $N$ system, we briefly consider the effect of the presence of a third particle ($N=3$ case). For simplicity we assume the particles to be located along a line ($x$ axis), and restrict the hydrodynamic kernel $G$ to nearest-neighbor interactions. By scaling times by  $\left(\frac{9}{512 \pi^2}f^2\tau\ell^4 / \eta^2 \right)^{-1}$ and numbering the particles along the positive $x$ axis, 
Eq.~\eqref{effectivebruit1} becomes
\begin{eqnarray}
\partial_t x_1&=& \left(\frac{\omega_1-\omega_2}{\omega_1+\omega_2}\right) \frac{1}{d_{1}^7},\,\, 
\partial_t x_3= \left(\frac{\omega_2-\omega_3}{\omega_2+\omega_3}\right) \frac{1}{d_{2}^7},
\\
\partial_t x_2&=& \partial_t x_3 + \partial_t x_1.
\end{eqnarray}
If we denote $d_1\equiv d_{12}= x_2-x_1$ and $d_2\equiv d_{23}=x_3-x_2$, it is easy to see that the  inter-particle distances obey 
\begin{eqnarray}\label{d3}
\partial_t d_1= \left(\frac{\omega_2-\omega_3}{\omega_2+\omega_3}\right) \frac{1}{d_{2}^7},\,
 \partial_t d_2= \left(\frac{\omega_2-\omega_1}{\omega_1+\omega_2}\right) \frac{1}{d_{1}^7}\cdot
\end{eqnarray}
Obviously the only interesting case is the one  where  the three intrinsic frequencies are all different. Without loss of generality we  can assume $\omega_1 < \omega_3$, and the only parameter is the relative magnitude of $\omega_2$ with respect to these two frequencies, so generically three different cases exist. If $\omega_2 < \omega_1 < \omega_3$ then the suspension  collapses (condenses) in finite time, and swims only on a finite length. If $\omega_1 < \omega_2 < \omega_3$ then one side (here $d_1$) collapses in finite time as $\sim (t_0-t)^{1/8}$ and the other side (here,  $d_2$) diverges (evaporates). If $\omega_1 < \omega_3 < \omega_2$ the whole system evaporates as $d_i\sim t^{1/8}$, leading to an average velocity on the order of $v\sim t^{-7/8}$, and swimming over an infinite distance $\Delta x = \int v dt \sim t^{1/8}$.
\subsection{Large groups of soft active particles: $N\gg1$}
The general equation describing the average motion of each soft active particle is Eq.~\eqref{effectivebruit1}. The right-hand side of this equation is the sum of the $N$ pairwise hydrodynamic interactions between the particles. Since all two-body interactions vanish when the relaxations times are identical ($\omega_i=\omega_j$), a suspension of identical soft active particle does not display any net motion when driven by incoherent active stresses. Any spatial distribution of the particles would thus correspond to a stationary state ($\partial_t\langle \rr_i\rangle=\bf 0$). Nonetheless, the ensemble of soft active particles should display collective diffusion, the dynamics of which could be in principle  obtained following the same framework as the one used above. 

Oppositely, suspensions of non-identical particles have in general a  non-zero collective velocity. If an ensemble of such particles is  randomly distributed in the plane, with the particles having a non-trivial (and 
well-behaved) distribution of relaxation time,  it follows from the central limit theorem that the collective velocity ${\bf V}\equiv\sum_i \langle\partial_t \rr_i\rangle/N$ has a random orientation, and a finite amplitude that scales as $V\sim1/\sqrt{N}$. In addition, the direction taken by such a ``swarm" of soft active particle would be dictated by the initial anisotropy of the particle distribution.    
\section{Coherent shaking}\label{coherent}

In this section, we  consider the scenario where we have a coherent driving. The active component of the force is assumed to be applied identically on all dimers, and to vary sinusoidally in time with prescribed frequency $\omega$, {\it i.e.} $f_i^{\rm act}(t)=f\cos(\omega t)$. This case is relevant to a wide range of experimental situations where a uniform external AC field is used to change the shape of responsive particles suspended in a viscous fluid. Examples include  magnetic fields used  to bend  arrays of magnetic colloidal filaments~\cite{goubault03}, 
 light or temperature cycling used to change the shape of  nematic elastomer colloids~\cite{buguin06}, or  electric fields used to deform emulsions of dielectric droplets.

Within the same linear response scheme as the one discussed in section~\ref{incoherent}, the deformations of the particles oscillate at the same frequency as the forcing, $\omega$, and are given by
\begin{equation}
\dl_i(t)=\delta\ell_i\cos(\omega t+\phi_i),
\label{lioscillations}
\end{equation}
where  the amplitude and phase are given by
\begin{equation}
\delta\ell_i = f \bigg|\sum_jR_{ij}(\omega)\bigg|,\quad 
\phi_i=\arg\left[\sum_jR_{ij} (\omega) \right]\cdot
\end{equation}

In order to compute the effective equation of motion, Eq.~\eqref{effective}, we need to calculate the time-averages 
\begin{equation}\label{toget}
\langle \delta\ell_i \delta\dot \ell_j\rangle =\frac{1}{2}\omega |\delta\ell_i||\delta\ell_j|\sin(\phi_i-\phi_j),
\end{equation}
and the  analysis differs between suspensions of identical or non-identical particles.

\subsection{Suspensions of non-identical particles}
We first consider the situation where  the soft particles have different relaxation times. In that case, and in the dilute limit, the response matrix $\bf R$ is dominated by its diagonal terms. Exploiting Eqs.~\eqref{Rii} and \eqref{Rij} it is straightforward to get
\begin{equation}\label{amp_pha}
|\delta\ell_i| = \frac{\mu_0f}{(\omega^2+\omega_i^2)^{1/2}}, \quad \phi_i= -\arctan\left(\frac{\omega}{\omega_j}\right),
\end{equation}
and therefore
\begin{eqnarray}
\langle \delta\ell_i \delta\dot \ell_j\rangle =\frac{\mu_0^2f^2}{2}\frac{\omega^2(\omega_i-\omega_j)}{(\omega^2+\omega_i^2)(\omega^2+\omega_j^2)},
\end{eqnarray}
so that the effective dynamics is fully described by
\begin{equation}
\label{effectivecoherent_hete}
\partial_t \langle {\bf r}_i \rangle = 
\frac{3 \ell^2f^2 \mu_0}{32}\frac{\omega^2(\omega_i-\omega_j)}{(\omega^2+\omega_i^2)(\omega^2+\omega_j^2)}\sum_{j\neq i}G'''(d_{ij}){\bf e}_{ij}.
\end{equation}
The result of Eq.~\eqref{effectivecoherent_hete} generalizes to $N$ particles the analysis done in Ref.~\cite{laugabartolo08} for $N=2$ dimers. We obtain an effective dynamics with a $\sim 1/d^4$ spatial decay. 
\subsection{Suspensions of identical particles}
We now focus on the case of an ensemble of identical soft particles. 
We assume that the particles all have the  same stiffness, $k$, and therefore identical relaxation frequency, $\omega_i\equiv\omega_0$.  The amplitudes in Eq.~\eqref{toget} are still dominated by the diagonal terms in $\bf R$ and we have 
\begin{equation}
|\delta\ell_i| = \frac{\mu_0f}{(\omega^2+\omega_0^2)^{1/2}}\cdot
\end{equation}
To get the phase differences, we have to look at the contribution of the off-diagonal terms in $\bf R$, as the leading-order term for $\phi_i$ given by Eq.~\eqref{amp_pha} predicts no phase difference for identical dimers. Exploiting the fact that we are in the dilute limit, and that therefore the phase differences are expected to be small ($\phi_i-\phi_j\ll1$), we can use Eqs.~\eqref{Rii} and \eqref{Rij} to explicitly obtain at leading order in $1/d_{ij}$
\begin{equation}
\phi_i-\phi_j= \frac{1}{\mu_0}\frac{\omega\omega_0}{\omega^2+\omega_0^2}\left[\sum_{k\neq i}\mu_{ik}-\sum_{k\neq j}\mu_{jk}
\right]\cdot
\end{equation}
The effective suspension dynamics is then obtained to be
\begin{eqnarray}
\partial_t\langle \rr_i\rangle&=&\frac{3\ell^2f^2}{32}\frac{\omega^2\omega_0}{(\omega^2+\omega_0^2)^2}\times \nonumber\\
&&\sum_{j\neq i}G'''(d_{ij})\left[\sum_{k\neq i}\mu_{ik}-\sum_{k\neq j}\mu_{jk}\right]\,{\bf e}_{ij},
\label{ACgeneral}
\end{eqnarray}
where the $\{\mu_{mn}\}$ are given by Eq.~\eqref{muij}. 

\subsection{General comments}
As a difference with the results of section~\ref{incoherent}, we see that in general, identical particles subject to AC forcing do display collective dynamics. Since only off-diagonal terms of $\bmu$ appear in Eq.~\eqref{ACgeneral}, the spatial decay of the velocity field scale generically as $G'''G''\sim 1/d^7$ in an unbounded fluid. We also note that Eq.~\eqref{ACgeneral}  actually represents a three-particle  interaction, with a non-pairwise additivity of the effective coupling. For a homogenous suspension with $N=2$ we have $\partial_t \langle x_1\rangle=\partial_t \langle x_2\rangle=0$, and no motion occurs on average, whereas $N=3$ particles do move (see below).  To provide with a physical interpretation of the results quantified by Eq.~\eqref{ACgeneral}, we now proceed to analyze two cases in detail.
\subsection{Shaking-induced crystallization} 
We first consider the dynamics of infinite one-dimensional chains. We assume that  the soft particles are all aligned along the $x$ direction, and number them such that $x_i < x_j $ if $i<j$. Such a setup may be achieved  experimentally by studying the dynamics of field-responsive particles  confined inside a capillary tube or in a micro-channel. 
As the $\mu_{mn}$ coefficients in Eq.~\eqref{ACgeneral} decay as $1/d^3$, we further make a nearest-neighbor coupling approximation and neglect any $\{i,i+n\}$ hydrodynamic coupling with $|n|>1$. Experimentally this assumption will be all the more valid that the system will be confined. Using the explicit expressions for the mobility coefficients, Eqs.~\eqref{muii} and \eqref{muij}, we then obtain the  form of the equations of motions for each dimer as
\begin{eqnarray}
\partial_t \langle x_i\rangle&=&\Lambda(\omega) 
G'''(d_i)
\left[G''(d_{i+1})-G''(d_{i-1})\right]\nonumber
\\
&+&\Lambda(\omega)G'''(d_{i-1})
\left[G''(d_{i})-G''(d_{i-2})\right]\label{effectiveAC},
\end{eqnarray}
where we have defined $d_i\equiv d_{i,i+1}$ and where the prefactor $\Lambda(\omega)$ is given by $\Lambda(\omega)=\frac{3}{128}\ell^4f^2\omega^2\omega_0/(\omega^2+\omega_0^2)^2$. Importantly, Eq.~\eqref{effectiveAC} shows that changing the forcing frequency does not  induce any qualitative changes in the dimers dynamics as it  can be absorbed into a rescaling of  time. We are thus left with a set of purely geometrical equations of motion. 

An obvious stationary state of Eq.~\eqref{effectiveAC} corresponds to a crystal order where all the distances are identical, $d_i=d,\,\forall i$. To investigate the stability of this ordered stated, we apply a small perturbation to the particle separations, $d_i=d+\delta d_i(t)$, and linearize the  equations of motion to get
\begin{equation}\label{linearized}
\partial_t \delta d_i=\Lambda(\omega)\left[G'''(d)\right]^2(\delta d_{i+2}-2\delta d_{i}+\delta d_{i-2}).
\end{equation} 
To compute the eigenvalues of Eq.~\eqref{linearized} we look for wave-like solutions of the form $\delta d_n(t)\equiv\delta d_n\exp [i\left (\Omega t+Qnd\right)]$, which leads to the dispersion relation of the compression waves along the chain of active dimers
\begin{equation}
i\Omega=-4\Lambda(\omega)\left[G'''(d)\right]^2\sin^2(Qd),
\label{dispersion}
\end{equation}  
with similar results  obtained if the nearest-neighbor approximation is relaxed. 
For all  shaking frequencies, the compression modes are overdamped, the crystal state is stable. In the low $Q$ (wavenumber) limit, the above dispersion relation is identical to the one of a chain of particles damped by a local drag force, and interacting through a harmonic potential. In other words the shaking induces an effective elasticity to this array  of $N$ non-interacting particles only dynamically coupled by the surrounding fluid flow. Note that such  mapping on the relaxation dynamics of a passive system is correct only for small perturbations around the steady state, and is in general not appropriate to describe the motion  of the dimers (see section \ref{pseudo}). 
\subsection{Small groups of soft active particles: Evaporation and spontaneous locomotion}
We now focus on finite-size one-dimensional chains (with $N >2$). In that case, the suspension does not posses a stationary state. 
This is best seen by writing down the evolution equation for the two end-particles, which are, under the nearest-neighbor approximation,
\begin{eqnarray}
\partial_{t} \langle x_1\rangle&=&\Lambda(\omega) G'''(d_1) G''(d_2), \\
\partial_{t} \langle x_N\rangle&=&- \Lambda(\omega) G'''(d_{N-1}) G''(d_{N-2}).
\end{eqnarray}
We see therefore that  $\partial_{t} \langle x_1\rangle< 0$ as well as $\partial_{t} \langle x_N\rangle > 0$, and therefore for any conformation of 
the chain, the velocity of the two end-particles  points outwards, The finite chain of active particles is thus seen to evaporate. 
Such evaporation takes place  in all finite suspensions, as confirmed by numerical simulations in two dimensions (not reproduced here).  

The evaporation dynamics can be analyzed in more detail in the case with $N=3$. In that situation, using the exact expression for $G(d)$ and rescaling  time to absorb all irrelevant prefactors, the three equations of motion reduce to
\begin{eqnarray}
\partial_t\langle x_1\rangle&=&-\frac{1}{d_1^{4}d_2^{3}}\label{dtx1}, \quad \partial_t\langle x_3\rangle=\frac{1}{d_2^{4}d_1^{3}}\label{dtx3},\\
\partial_t\langle x_2\rangle&=& \partial_t\langle x_1\rangle + \partial_t\langle x_3\rangle.
\end{eqnarray}
It is then straightforward to show that $(\partial_t d_1)/d_1=(\partial_t d_2)/d_2$ and thus that the ratio between the two distances, $\alpha =d_2/d_1$, is a constant of the motion. It then follows that the inter-particle distances grows with time as
$d_1(t)\sim\alpha^{-1/2}t^{1/8}$ and   $d_2(t)\sim\alpha^{1/2}t^{1/8}$. 
Hence the growth speed of the suspension varies as $\partial_t(d_1+d_2)\sim(\alpha^{1/2}+\alpha^{-1/2})t^{1/8}$. 
Let us now consider  the average speed, $v=\frac{1}{3}(\partial_t\langle x_1\rangle+\partial_t\langle x_2\rangle+\partial_t\langle x_3\rangle)$, of the three dimers. It is easy to see that  $v=\frac{2}{3}(1-\alpha)\partial_t d_1$. For a symmetric conformation, $d_1=d_2$, and the system does not swim on average. However, the hydrodynamic coupling between the dimers induces a spontaneous left-right symmetry breaking. Indeed, an infinitesimal perturbation to the ratio, say $d_2/d_1=1+\epsilon$,  cannot relax as $d_2/d_1$ is a constant of motion. This small swarm of active particles acquires therefore a spontaneous net motion with a velocity decreasing with time as $v\sim t^{-7/8}$,  implying that the center of the suspension  can travel over an infinite distance, $\Delta x \sim t^{1/8}$. Three identical active soft particles swim therefore cooperatively. 
Note that the dilatation width, $\partial_t(\langle d_1+d_2\rangle)$,  and the swimming speed,  $v$, decrease with the same scaling, and $\partial_t(\langle d_1+d_2\rangle)/|v|=3(1+\alpha)/2|1-\alpha| > 1$, which implies that the system evaporates faster than that it swims for any initial conformations. 
\subsection{Large  groups of soft active particles}
Within the approximations made in this paper, the collective mode locomotion discussed above in the case of $N=3$ can only exist in the case of finite suspensions. Indeed, from Eq.~\eqref{effectiveAC} it is straightforward to show that for $N$ particles, the average speed of the suspension, $v$, scales as 
 $v\sim\frac{1}{N}\sum_i(d_i-d_{i+1})/(d_i^4d_{i+1}^4)$ up to ${\cal O}({1}/{N})$ boundary terms. So, if the $d_i$ are finites, we have the inequality $\frac{1}{N}|d_N-d_1|/\max(d_i)^8<v<\frac{1}{N}|d_N-d_1|/\min(d_i)^8$. As a result,  for all spatial distributions of the elastic dimers, the collective locomotion speed if at most of order ${\cal O}(1/N)$. 
  Whereas spontaneous motion  takes place for small groups of soft active particles, the effect  vanishes therefore in the thermodynamic limit. Note however that  a non-zero collective locomotion could arise from nonlinear effects in the conformational dynamics of the particles. This non-trivial extension lies beyond the scope of this paper.
\section{Conclusion}
We now conclude by returning to the initial question that motivated this work: Does the hydrodynamic interplay between the shape fluctuations and the positions of soft active particles yield unanticipated collective behavior? As we demonstrated in this paper, the answer is yes. On the basis of a minimal model for  hydrodynamically coupled soft dimers, we showed that  despite the fact that these particles are intrinsically non-motile, we obtained the following  results for their collective dynamics. Finite ensembles of non-identical soft  particles experience collective swimming when driven by an uncorrelated stochastic internal forcing. The direction and the amplitude of the resulting collective swimming is set by the initial anisotropy of the particles positions. In addition, these ``swarms" condense or evaporate depending on the distribution of relaxation times in the suspension. {  In contrast,  soft  particles having identical relaxation times, can only  display collective diffusion.}
 When shaken by a monochromatic and homogeneous field, finite ensemble of identical 
soft active particles always  display collective swimming and evaporation in the dilute limit for all initial positions.
In the large $N$ limit,  the suspension displays shaking-induced crystallization. 
However, in this large $N$ limit, the collective swimming velocity   vanishes algebraically due to the fast decay of the correlation of the shape oscillations ($1/d^3$).

The cancelation of the collective swimming speed in the thermodynamic limit might result from the linear response  approximation we used to compute the shape fluctuations of the particles. {In the dual and simpler problem of collective pumping,  it has been recently 
shown that an array of hydrodynamically-coupled filaments  driven with the same constant torque can be unstable, resulting in the propagation of metachronal waves, and fluid pumping, see e.g.~\cite{lenz08,joanny07}. A similar rectification phenomena could in principe exist in suspensions of freely moving soft active particles. This intrinsically non-linear effect would likely result in a motility transition in suspensions of reversibly actuated particles.}

\section*{Acknowledgements}
This research was funded in part by the NSF (grants CTS-0624830 and CBET-0746285 to EL). We also acknowledge funding from a PEPS-PTI fellowship from CNRS.

\bibliography{dimersIIbib}
\end{document}